\documentclass[aps,pre,twocolumn]{revtex4-1}
\usepackage{amsmath}
\usepackage{amsfonts}
\usepackage{amssymb}
\usepackage{graphicx}
\usepackage[colorlinks,linkcolor=blue,anchorcolor=blue,citecolor=blue,urlcolor=black]%
{hyperref}
\usepackage{mathrsfs}
\usepackage{dcolumn}
\usepackage{bm}
\usepackage{epsfig}
\usepackage[version=3]{mhchem}%
\def\be{\begin{equation}}
\def\ee{\end{equation}}
\begin{document}
\title{Statistics of Chaotic Resonances in an Optical Microcavity}
\author{Li Wang$^{1}$}
\author{Domenico Lippolis$^{2,3}$}
\author{Ze-Yang Li$^{1}$}
\author{Xue-Feng Jiang$^{1}$}
\author{Qihuang Gong$^{1}$}
\author{Yun-Feng Xiao$^{1}$}
\email{yfxiao@pku.edu.cn}
\altaffiliation{URL: www.phy.pku.edu.cn/$\sim$yfxiao/}

\affiliation{$^{1}$State Key Laboratory for Mesoscopic Physics and School of Physics,
Peking University; Collaborative Innovation Center of Quantum Matter, Beijing
100871, China}
\affiliation{$^{2}$Institute for Advanced Study, Tsinghua University, Beijing 100084, China}
\affiliation{$^{3}$Faculty of Science, Jiangsu University, Zhenjiang 212013, China}
\date{\today}

\begin{abstract}

Distributions of eigenmodes are widely concerned
in both bounded and open systems.
In the realm of chaos, counting resonances can characterize
the underlying dynamics (regular vs. chaotic), and is
often instrumental to identify classical-to-quantum correspondence.
Here, we study, both theoretically and experimentally, the statistics of chaotic resonances in an optical
microcavity with a mixed phase space of both regular and chaotic dynamics. Information on the
number of chaotic modes is extracted by counting regular modes, which couple to the former via
dynamical tunneling. The experimental data are in agreement with a known semiclassical prediction
for the dependence of the number of chaotic resonances on the number of open channels, while they
deviate significantly from a purely random-matrix-theory-based treatment, in general. We ascribe
this result to the ballistic decay of the rays, which occurs within Ehrenfest time, and importantly,
within the timescale of transient chaos. The present approach may provide a general tool for the
statistical analysis of chaotic resonances in open systems.

\end{abstract}

\maketitle

\emph{Introduction.}  The statistics of chaotic resonances has been a central topic of theoretical
and experimental interest for
decades \cite{Wirzba, folk96, ElastRes96, Weidenm, ColdAtomRes, CondMatRes, CaoWier, PLS00} , as a doorway to better understand chaotic scattering in quantum mechanics \cite{BR,Gasp89,Anlage,squid}.
Counting chaotic resonances notably finds applications
in optical resonators \cite{Star00,Hack05,SWM}, where chaos can be used, for
example, to enhance energy storage \cite{storage13} or enhance coupling
\cite{Yang_dyntun,xiao13}.
In the realm of fundamental problems, an estimate of the number of states
within a certain frequency interval is given by the fractal Weyl law
\cite{LSZ,ketzWeyl,IShudoSchom,Zwor_pre,Zwor_prl}. Predicted more than a decade ago, this theory is
still awaiting experimental confirmation at optical frequencies.
The delay is mainly due to two obstacles: (i) the theoretical problem of
accounting for partial absorption~\cite{Nonnenmacher,Schonwetter}; and (ii) the experimental challenge
of analyzing overlapping resonances~\cite{WierMain}.
By introducing methods to overcome these hurdles, we
theoretically and experimentally study the statistics of
chaotic resonances in two-dimensional asymmetric optical microcavities, and thus present a
significant result in the context of fractal Weyl laws at optical frequencies.
An absorber is placed underneath the microcavity appositely to
realize a virtually full opening, and therefore solve the problem of partial absorption.
In order to handle overlapping resonances,
we exploit the mixed classical phase space of the present cavity,
and exclusively count high-$Q$ regular modes,
easily recognized in the
measured spectra due to their narrow linewidths, and that enables us to draw
information on the low-$Q$ chaotic modes, coupled to the regular ones via
dynamical tunneling~\cite{davis81}.

\begin{figure}[tb]
\centerline{\includegraphics[width=8cm]{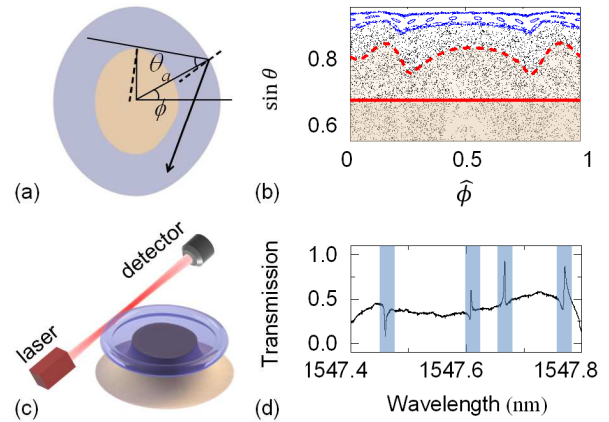}}
\caption{(color online) (a) Sketch of the deformed microcavity with an inner absorber,
characterized by the angle $\theta_{a}$. (b) Poincar\'{e} surface of section
of the microcavity ($\hat\phi\equiv\phi/2\pi$) with deformation factor
$\eta=11.7\%$. Blue orbits at the top are regular, black ones are chaotic.
The red solid line indicates the angle of total internal
reflection, while the dashed curve is given by an absorption angle such that
$r\simeq0.77$. Different shades of color indicate loss to the absorber
(lighter) and by refraction into air (darker). (c) Schematic representation of the free-space coupled cavity
system. (d) A typical transmission spectrum with the high-$Q$ regular
modes highlighted.}%
\label{figureone}%
\end{figure}

\emph{Theoretical model.} The dynamics inside the cavity is described by a non-Hermitian Hamiltonian
$H=H_{0}+V$, where $H_{0}$ has eigenstates $|C_{\omega}\rangle$ (regular) and
$|C_{n}\rangle$ (chaotic), while $V$ represents the coupling between them
\cite{BTU,Backer1}.
Following a standard approach \cite{Fano,AnYang}, the
electromagnetic field excited by the incident beam can be written as a
superposition of one regular and several chaotic modes, $\psi(x,t)=a_{\omega
}(t)c_{\omega}(x)e^{-i\omega t}+\sum_{n}b_{n}(t)c_{n}(x)e^{-i\omega_{n} t}$.
The $a_{\omega}$ and $b_{n}$ are oscillator amplitudes driven by the laser
beam, coupled to each other with strength $V_{n}$ (real), with damping rates
$\gamma_{\omega}$ and $\gamma_{n}$, respectively. Then, in the overdamped
regime ($|V_{n}|\ll\gamma_{n}$), the dynamical equations for the
time-dependent envelopes take the form \cite{Yang_dyntun}
\begin{subequations}
\begin{align}
\dot{b}_{n}+\gamma_{n}b_{n}  &  =f_{n}E_{0}-V_{n}a_{\omega},\\
\dot{a}_{\omega}+\left[  \gamma_{\omega}+i(\omega_{0}-\omega)\right]
a_{\omega}  &  ={\textstyle\sum\limits_{n}}V_{n}b_{n}, \label{coup_osc}%
\end{align}
\end{subequations}
where $f_{n}$ is the coupling strength of the $n$-th chaotic mode with the
laser beam of amplitude $E_{0}$ and frequency $\omega_{0}$. We are interested
in the steady-state solution,
obtained by setting $\dot{a}_{\omega}=\dot
{b}_{n}=0$.
The amplitude $a_\omega$ of the envelope of the regular mode
is then found to be
\begin{equation}
a_{\omega}=\frac{E_{0}\sum\limits_{n}f_{n}\frac{V_{n}}{\gamma_{n}}}{\left[
\gamma_{\omega}+i(\omega-\omega_{0})\right]  +\sum\limits_{n}\frac{V_{n}^{2}%
}{\gamma_{n}}}. \label{excit_amp}%
\end{equation}
We only sum over $n_{\gamma}$ chaotic modes with
linewidth smaller than a certain value set by $\gamma$,
and use the averages $\bar{f},\bar{V},\bar{\gamma}$ to approximate the
summations as ${\textstyle\sum_{n}}f_{n}\frac{V_{n}}{\gamma_{n}}\sim
n_{\gamma}\frac{\bar{f}\bar{V}}{\bar{\gamma}},{\textstyle\sum_{n}}\frac
{V_{n}^{2}}{\gamma_{n}}\sim n_{\gamma}\frac{\bar{V}^{2}}{\bar{\gamma}}$
\cite{SI}. The result is, at resonance ($\omega=\omega_{0}$),
\begin{equation}
|a_{\omega}|^{2}=\epsilon^{2}\frac{n_{\gamma}^{2}}{\Gamma^{2}+n_{\gamma}^{2}},
\label{ex_prob_3par}%
\end{equation}
where $\epsilon=E_{0}\bar{f}/\bar{V}$, and $\Gamma=\gamma_{\omega}\bar{\gamma
}/\bar{V}^{2}$.

Equation~(\ref{ex_prob_3par}) is central to the present work, as it bridges
between the number of excited regular modes (proportional to $|a_{\omega}%
|^{2}$), which we measure directly, and the number of chaotic modes $n_{\gamma
}$, which we estimate. In the spirit of Weyl law, we test
the prediction \cite{SchomTwor}
\begin{equation}
n_{\gamma,\mathrm{Weyl}}= MN^{-1/\mu\tau_{d}}\left[  1-\frac{1}{\tau_{d}}%
\frac{1}{1-e^{-\gamma}}\right]  , \label{ngam_law}%
\end{equation}
for the dependence of $n_{\gamma}$ on the number of open channels $N$. Here
$M$ is the total number of chaotic states, $\tau_{d}=M/N$ is the mean dwelling
time of a ray in the cavity, and $\mu$ is of the order of the Lyapunov
exponent of the chaotic dynamics \cite{SI}.
We focus on the role of the prefactor $N^{-1/\mu\tau_{d}}$, which accounts for
the instantaneous-decay modes escaping from the system within the Ehrenfest
time of quantum-to-classical correspondence \cite{SchomJacq}. By removing the
prefactor, the remainder of Eq.~(\ref{ngam_law}) is solely based on random
matrix theory (RMT) \cite{ZycSom}, and will also be tested against the
observations.

\emph{Absorber in the optical cavity.}  In order to achieve the full opening required to test these predictions, we
introduce an absorber in the cavity.
In the analysis,
the dielectric microcavity [Fig.~\ref{figureone}(a)] has the deformed circle
$\rho(\phi)$ as boundary \cite{SI},
and encloses an absorber of shape $\rho(\phi)-R$. Figure~\ref{figureone}(b) shows
the classical phase space, together with the critical line of total internal
reflection ($\sin\theta_{c}$), as well as the line given by the incidence
angle $\theta_{a}$, below which the reflected ray hits the absorber
\cite{ThetaNote}.
We assume that the rays hitting the absorber are completely absorbed. 
The rays that escape the cavity by refraction into the air with an angle of
incidence $\theta\ll\theta_{c}$ are very lossy, 
and, as such, they
are not expected to contribute to the excitation of the regular modes,
consistently with Eqs.~(\ref{excit_amp}) and~(\ref{ex_prob_3par}).

For that reason, we only take into account the states supported on a strip of
the chaotic phase space with momentum above a certain threshold, $\sin
\theta>\sin\theta_{\mathrm{th}}$, to be chosen below but close enough to the
critical line of total internal reflection. Let us introduce the notation
$\xi\equiv\sin\theta_{a}-\sin\theta_{\mathrm{th}}$ to indicate the strip of
the phase space opened by the absorber. Using the picture of
Ref.~\cite{SchomTwor}, there are effectively $N$ open channels out of the $M$
Planck cells available in the phase space, produced by the absorber (full
opening, $N_{a}$) and the refraction out of the cavity (partial opening,
$N_{r}$), so that the mean dwelling time of a ray is given by
\begin{equation}
\tau_{d}=\frac{M}{N_{a}+N_{r}},
\label{T}%
\end{equation}
with $N_{r}=\frac{M}{A}\int_{\sin\theta_{a}}^{\sin\theta_{c}}d\sin\theta T(\sin
\theta)$, $T$ transmission coefficient according to Fresnel law, and
$A$ area of the phase space in exam, while $N_{a}=M\xi/A$.
\begin{figure}[tb]
\includegraphics[width=8.5cm]{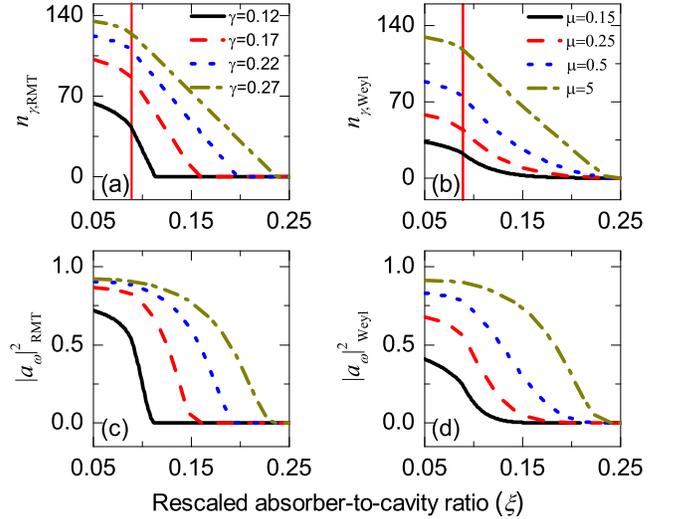}
\caption{(color online) (a), (b) Number of chaotic states $n_{\gamma}$ vs. the
rescaled absorber-to-cavity ratio $\xi$, obtained by RMT and Weyl law,
respectively. (c), (d) The corresponding expectations for $|a_{\omega}%
|^{2}$.
In (d), the smaller $\mu$, the more visible the tail of the
excitation probability. The red vertical line corresponds to the critical
angle, $\sin\theta_{c}\simeq0.69$. Here $\sin\theta_{\mathrm{th}}=0.6$.}
\label{figuretwo}%
\end{figure}

Recalling the original purpose of studying the statistics of chaotic
resonances, we proceed by steps and first examine the RMT-based prediction
$n_{\gamma,RMT}= M\left[  1-\frac{1}{\tau_{d}}\frac{1}{1-e^{-\gamma}%
}\right]  $, rewritten as (set $\hat{\xi}=1/\tau_d=\xi/A+N_{r}/M$)
\begin{equation}
n_{\gamma,RMT}=M\left[  1-\frac{\hat{\xi}}{1-e^{-\gamma}}\right]  .
\label{RMT_x}%
\end{equation}
The theoretical expectation is portrayed in Fig.~\ref{figuretwo}(a):
$n_{\gamma,\mathrm{RMT}}$ changes relatively slowly for $\xi$ small such that
$\theta_{a}<\theta_{c}$, when the loss is mainly due to refraction into air.
Otherwise $n_{\gamma,\mathrm{RMT}}$ decreases more rapidly and linearly with
$\xi$ in the region of total internal reflection, when the loss is entirely
due to the absorber. Plugging
Eq.~(\ref{RMT_x}) into Eq.~(\ref{ex_prob_3par}),
the probability of excitation
of the high-$Q$ regular modes $|a_{\omega}|^{2}_{\mathrm{RMT}}$ starts to fall
off as $\xi$ reaches some critical value, controlled by the parameter $\gamma$
[Fig.~\ref{figuretwo}(c)]. The other parameter $\tilde{\Gamma}=\gamma_{\omega}\bar{\gamma}/M\bar{V}^2$
controls the slope of the curve. It is noted that the
probability of excitation of the regular modes decreases dramatically
when the
loss is entirely due to the absorber, in which case the system is
fully open.

On the other hand, the semiclassical estimate~(\ref{ngam_law}) becomes, as a
function of~$\hat{\xi}$,
\begin{equation}
n_{\gamma,\mathrm{Weyl}}=\frac{M^{1-\hat{\xi}/\mu}}{\hat{\xi}^{\hat{\xi}/\mu}
}\left[  1-\frac{\hat{\xi}}{1-e^{-\gamma}}\right]  . \label{ngam_law_x}%
\end{equation}
The quantity $\mu$, of the order of the Lyapunov exponent of the chaotic
region of the phase space \cite{SI},
 is what really
characterizes~(\ref{ngam_law_x}), 
which resembles the linear RMT prediction~(\ref{RMT_x})  for large enough $\mu$, and otherwise
becomes visibly nonlinear [Figs.~\ref{figuretwo}(a) and \ref{figuretwo}(b)] when $\mu\ll1$.
This nonlinearity produces a \textit{characteristic tail} in the curve
expressed by Eq.~(\ref{ex_prob_3par}) [Fig.~\ref{figuretwo}(d)], meaning that
the effect of the Ehrenfest time scale on the excitation of the regular modes
is most evident slightly above the onset of chaos.

\begin{figure}[ptb]
\includegraphics[width=8.5cm]{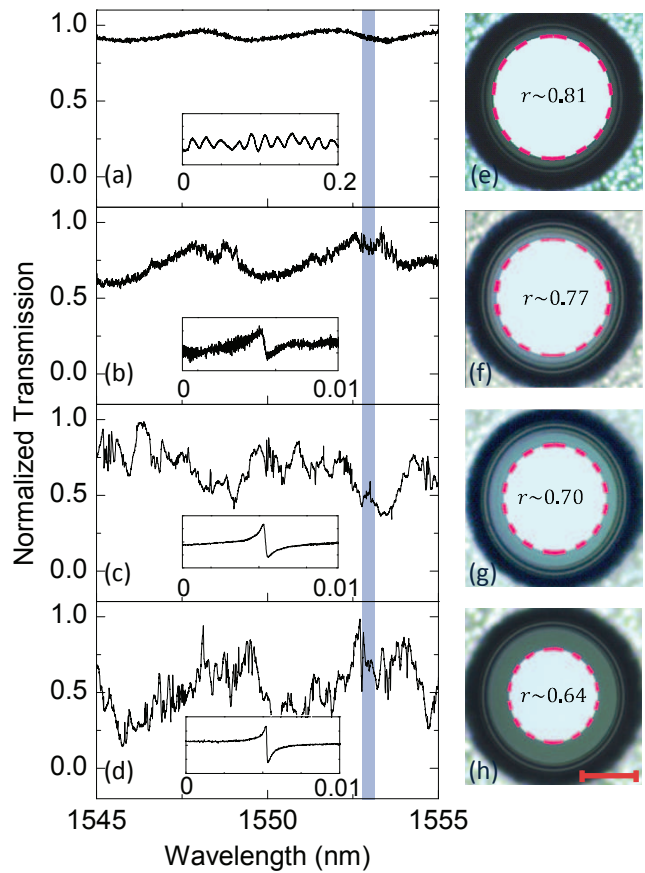}
\caption{(color online) Normalized
transmission and top-view optical images of the cavity with $r\simeq0.81$ [(a)
and (e)], $0.77$ [(b) and  (f)], $0.70$ [(c) and (g)], and $0.64$ [(d) and (h)]. Inset of (a) shows background noise. Insets of (b)-(d) show the high-$Q$
modes. Reflection of the silica-to-silicon interface results in a brighter
color for the silicon pillar in the optical image (boundary shown by red
dashed curves). Scale bar is $50$ $\mathrm{\mu}$\textrm{m}. }%
\label{figurethree}%
\end{figure}

\emph{Experimental setup and measurement.} The experimental apparatus consists of a deformed toroidal microcavity
\cite{xiao13} coupled to a laser beam (of wavelength $\lambda\simeq1550$
or $635$ \textrm{nm}), as shown in Fig.~\ref{figureone}(c). The
microtoroid (refractive index $\simeq1.44,1.46$ depending on
$\lambda$) has principal (minor) diameters of $120$ $\mathrm{\mu}$\textrm{m}($5$ $\mathrm{\mu}$\textrm{m}),
consistently with the two-dimensional model \cite{SI}.
Thus the effective
Planck constant $h_{\mathrm{eff}}\sim\lambda/a\sim10^{-2}$ ($a$: principal
diameter) justifies the semiclassical analysis.  
The microcavity is fabricated through optical lithography, buffered \ce{HF} wet etching,
\ce{XeF2} gas etching, and \ce{CO2} pulse laser irradiation. The resulting
silica microtoroid is supported by a silicon pillar of similar shape, which has a high refractive index ($\simeq3.48,3.88$),
and it acts as the absorber in the model.
After each measurement of the free-space transmission spectrum [Fig.~\ref{figureone}(d)],
the top diameter of the silicon pillar, connected with the silica disk,
is progressively reduced by a new isotropic \ce{XeF2} dry etching process.
In this way we control the openness of the
microcavity with the ratio $r$ between the top diameters of pillar and toroid.
 Finite element method simulations show that the light power
decreases to less than $5\%$ of the input value, when propagating by a
distance of $20$ $\mathrm{\mu}$\textrm{m} inside the $2$-$\mathrm{\mu}%
$\textrm{m}-thick silica waveguide bonding with a silicon wafer \cite{SI},
as is reasonable to expect, given the high refractive index of the silicon.
Thus the silicon pillar acts as a full absorber,
consistently with the present model.
On the other hand, high-$Q$ regular modes
living inside the toroidal part, whose cross section has minor diameter of $5$
$\mathrm{\mu}$\textrm{m}, do not leak into the silicon pillar and therefore
are not directly affected by the pillar size \cite{SI}.
The dependence of the
free-space transmission spectra on the pillar size is shown in
Fig.~\ref{figurethree}. When the pillar approaches the inner edge of the
toroid [Figs.~\ref{figurethree}(a) and \ref{figurethree}(e), $r\simeq0.81$], no high-$Q$
regular modes are observed in the spectrum, since most of the
probe laser field in the cavity radiates into the silicon and cannot tunnel to
high-$Q$ regular modes. As we gradually reduce the size of the pillar
[Fig.~\ref{figurethree}(f), $r\simeq0.77$], increasingly many high-$Q$ modes
appear in the spectrum [Fig.~\ref{figurethree}(b)]. When the
absorber-to-cavity ratio $r$ is small enough [Figs.~\ref{figurethree}(g) and
\ref{figurethree}(h), $r\lesssim0.7$], the transmission no longer changes sensibly
[Figs.~\ref{figurethree}(c) and \ref{figurethree}(d)], and the number of high-$Q$ modes in the
spectrum also stabilizes.

\begin{figure}[ptb]
\includegraphics[width=8cm]{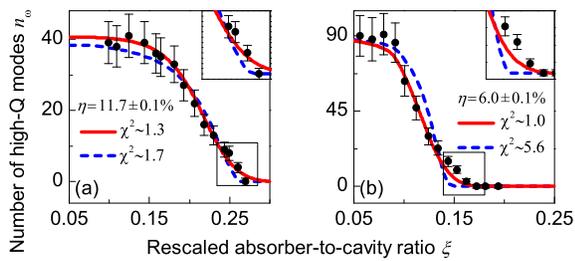}
\caption{(color online) (a), (b) Number of high-$Q$ regular modes ($n_{\omega}$)
observed experimentally in the microcavities of $\eta=11.7\%$, $6\%$ respectively, vs.
rescaled absorber-to-cavity ratio $\xi$. Blue dashed and red solid curves
are respectively RMT and Weyl-law best fits, whose goodness is assessed by $\chi^{2}$. Here $\sin\theta_{c}\simeq0.69$
and $\sin\theta_{\mathrm{th}}=0.6$. Inset: area where the two curves
differ most.}%
\label{figurefour}%
\end{figure}

\emph{Statistics of chaotic resonances.}
As anticipated, we use the transmission spectra to test the theory, by
counting the excited high-$Q$ regular modes for different sizes of the silicon
pillar \cite{Note}.
 The results are illustrated in Fig.~\ref{figurefour}, for two microcavities of
distinct deformations. 
In the RMT-based approach
[Eqs.~(\ref{ex_prob_3par}) and~(\ref{RMT_x})]
we have two
fitting parameters, $\gamma$ and $\tilde{\Gamma}$, while the total number of
chaotic states is estimated theoretically as $M\simeq A/h_{\mathrm{eff}}$ ($A$
area of the phase space we consider).
Figure~\ref{figurefour}(a) shows overall
agreement between the experimental data (dots) and this theory (blue dashed curve),
which however deviates from the tail visible at larger sizes of the absorber.
The discrepancy becomes more apparent for the cavity with a lower deformation factor $\eta$ [Fig.~\ref{figurefour}(b)].
A quantitative test of the goodness of the fit yields a reduced $\chi^{2}%
\simeq1.7,5.6$ for $\eta\simeq11.7\%,6\%$ respectively \cite{chi-s,SI},
the latter deviating significantly from
the optimal value of unity.
The fitted maximum escape rate $\gamma\simeq0.2,0.3$
is the inverse (in units of Poincar\'e time) of the minimum escape time
$\tau_{esc}$
of the chaotic rays contributing to the excitation of the regular modes,
from which $Q=2\pi\nu\tau_{esc}\sim10^{3}$, on average ($\nu$ is the frequency of the laser beam).
We find this estimate consistent with the typical order of $Q$ independently
obtained from ray-dynamics simulations \cite{SI},
which suggests the fitted
parameter makes physical sense.

Next, we test the semiclassical correction~(\ref{ngam_law_x}), using the finite time Lyapunov exponent $\mu$ evaluated by direct iteration \cite{finLyap,SI},
with the estimated parameter $M$ and the fitted parameters $\gamma$ and $\Gamma$.
It is found that the semiclassical correction (red solid curves) fits the experimental data better
than the RMT-based estimate, especially at the smaller deformation,
where the two predictions differ the most due to the smaller $\mu$ [cf. Fig. \ref{figuretwo}]. Here
$\chi^{2}\simeq1.3,1$, indicating good agreement. In particular, we are now able
to account for the tail of the curve, which corresponds to the microcavity
having the largest openings and thus with the maximum number of instantaneous
decay states, where the
semiclassical correction is decisive.
More experimental and fitting results at different wavelengths and deformations support the above explanation \cite{SI}.

\emph{Conclusion and discussion.} Let us summarize the work done and the
results obtained. By counting the high-$Q$ regular modes excited via dynamical tunneling as a function of the number of open
channels, we have studied the statistics of the chaotic resonances in a dielectric microcavity
with a full absorber.

As main result,
the experimental data deviate from a purely RMT-based prediction, while they
exhibit better agreement with a semiclassical expression that factors out
the number of instantaneous decay modes.
Importantly, the latter estimate depends on the Lyapunov exponent of the
chaotic dynamics, and it accounts for
a characteristic tail in the decay of the number of regular modes, which we
interpret as a signature of the ballistic escape of the rays into the
absorber, occurring within Ehrenfest time.

\begin{figure}[ptb]
\includegraphics[width=8cm]{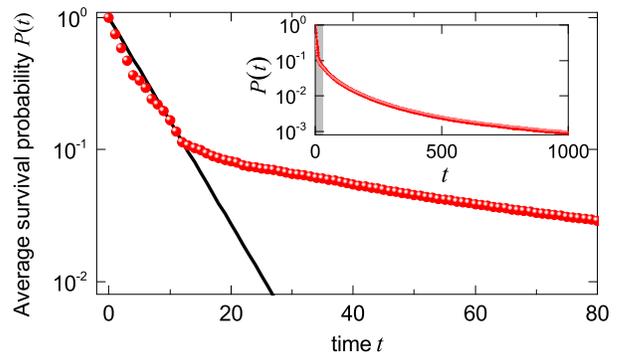}
\caption{(color online) Survival probability in the chaotic region (logarithmic scale). Points:
average survival probability $P(t)$ of a ray in the
microcavity vs. $t$ (in units of Poincar\'e time) at $\eta=11.7\%,\xi=0.13$, from $10^{6}$ randomly-started trajectories.
Line: $P(t)=\mathrm{exp}(-t/\tau_{d})$, $\tau_{d}=6$. Inset: the
long-time simulation showing algebraic decay.}%
\label{figurefive}%
\end{figure}

Although the theoretical analysis does not take into account either partial transport
barriers~\cite{MeissPBs,KetzPBs}, or the "sticky" dynamics at the regular-chaotic border~\cite{ketzWeyl,IShudoSchom}, which all
cause long-time correlations to decay algebraically rather than
exponentially,
we argue in what follows that the current model is suitable for
the resolution of
 our experiment. Figure~\ref{figurefive} illustrates
the survival probability in the chaotic region, obtained from extensive ray-dynamics simulations of
the microcavity-shaped billiard:  
despite an overall power-law decay, a closer look at the short-time dynamics reveals that
the decay is initially exponential, behavior known as transient
chaos~\cite{LaiTel}.

The estimates for the dwelling time $\tau_{d}$, and the fitted value from the data of
$\gamma$, maximum decay rate of the chaotic resonances, are both within the
timescale of exponential decay. That suggests that the suppression by the
absorber of the longest-lived resonances (that scale algebraically with $\gamma$ \cite{IShudoSchom}) alone
does not affect the number of excited WGMs measured
in the experiment, and therefore is not detected by the current apparatus.

\emph{Acknowledgments---}
This project was supported by the 973 program (Grant No. 2013CB921904 and No.
2013CB328704) and the NSFC (Grants No. 61435001 and No. 11474011). D.L.
acknowledges support from NSFC (Grant No. 11450110057-041323001). Z.-Y.L. was
supported by the National Fund for Fostering Talents of Basic Science (Grant
No. J1103205). L.W. fabricated the microcavities and performed measurements and numerical simulations. D.L.
developed the theoretical model. Y.-F.X. supervised and coordinated the project.
All authors contributed to the discussions and wrote the manuscript.
%\end{acknowledgments}

\clearpage
\begin{widetext}
\setcounter{section}{0}
\setcounter{equation}{0}
\setcounter{figure}{0}
\setcounter{page}{1}
\ 

{\center{\bf \large Supplemental Information to `Statistics of chaotic resonances in an optical microcavity'}}
\ \\

This Supplementary Material is organized as follows. In Sec. I, we describe the boundary shape of the
cavity and its unidirectional emission. In Sec. II, the $Q$ factor magnitude distribution of the chaotic rays
is estimated by ray-dynamics simulation. In Sec. III, we show geometry and field distribution of microtoroid
by SEM images and finite element method modeling. In Sec. IV, power loss in the
silica waveguide bonding with a thick silicon layer is studied. In Sec. V, we
provide a cross-check that many high-$Q$ modes do exist in the deformed microcavity by exciting
them directly through a tapered fiber. In Sec.VI, the process of obtaining the number of high-$Q$ regular modes
is described. Sec. VII deals
with the definition of the Ehrenfest time and rescaling of Lyapunov exponent in the Wey law. In Sec. VIII,
we explain the $\chi^{2}$ test used to assess the goodness of our fits. In Sec. IX, more experimental and fitting
results are shown. Sec. X contains tables with the quantities and parameters used in our final results. In Sec. XI,
the phase space with a small deformation factor together with the survival probability
of a ray in the chaotic region are shown. In Sec. XII, we study the relation between the decay rates of WGMs and their coupling to
chaotic modes.

\ 

\section{Boundary shape of the cavity}

The deformed microtoroid cavity in our experiment has a boundary shape given
by the curve
\begin{equation}
\rho(\phi)=\left\{
\begin{array}
[c]{cc}%
\rho_{0}(1+\epsilon\sum_{i=2,3}a_{i}\cos^{i}\phi) & \mathrm{for}\cos\phi
\geq0,\\
\rho_{0}(1+\epsilon\sum_{i=2,3}b_{i}\cos^{i}\phi) & \mathrm{for}\cos\phi<0,
\end{array}
\right. \label{df_circ}%
\end{equation}
with $a_{2}=-0.1329,a_{3}=0.0948,b_{2}=-0.0642$, and $b_{3}=-0.0224$. The WGMs in
the deformed microcavity have been demonstrated to possess ultrahigh quality
factors in excess of $10^{8}$ in the $1550$ nm wavelength band and to exhibit
highly directional emission towards the $180^{\circ}$ far-field direction,
which emits tangentially along the cavity boundaries at polar angles $\phi
=\pi/2$ and $\phi=3\pi/2$. The deformation is controlled by $\eta=(d_{\max
}-d_{\min})/d_{\max}$, $d_{\max}$ and $d_{\min}$, respectively, the maximum and
minimum diameters of the cavity. The parameter $\eta$ is related to $\epsilon$
through $\eta=\epsilon\left\vert a_{2}+a_{3}+b_{2}-b_{3}\right\vert /2$. The $\eta$ of
cavity shape we used to draw the Poincar\'{e} surface of section in the main text
is $11.7\%$.

\section{Estimating the $Q$ factor magnitude distribution of the chaotic rays}

\begin{figure}[tbh]
\centerline{
\includegraphics[width=.4\textwidth]{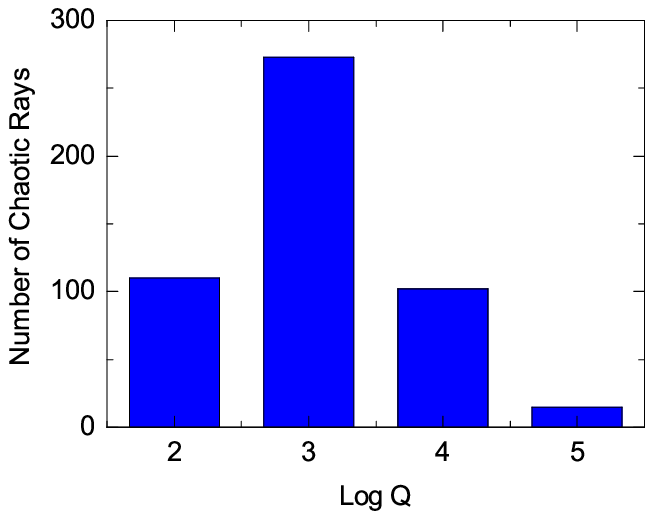}}\caption{Statistics of
the quality factor for the chaotic rays in the deformed microcavity, from a
ray-dynamics simulation.}%
\label{Qstat}%
\end{figure}A ray-dynamics simulation is performed on the billiard with
boundary set by Eq.~\ref{df_circ} ($\epsilon$=1). A swarm of initial points,
randomly chosen in the chaotic region of the phase space, is iterated $1000$
times. Each ray is considered to have escaped when it falls below the critical
line of total internal reflection, and its quality factor $Q$ is measured as
$Q=2\pi\tau\nu$, with $\tau$ lifetime of the ray and $\nu$ frequency of the
light. The statistics shown in Fig.~\ref{Qstat} indicates that $Q$ is mostly
of the order of $10^{3}$, confirming the estimation from the experimental results.

\section{Geometry and field distribution of the microtoroid}

\begin{figure}[tbh]
\centerline{
\includegraphics[width=.5\textwidth,clip]{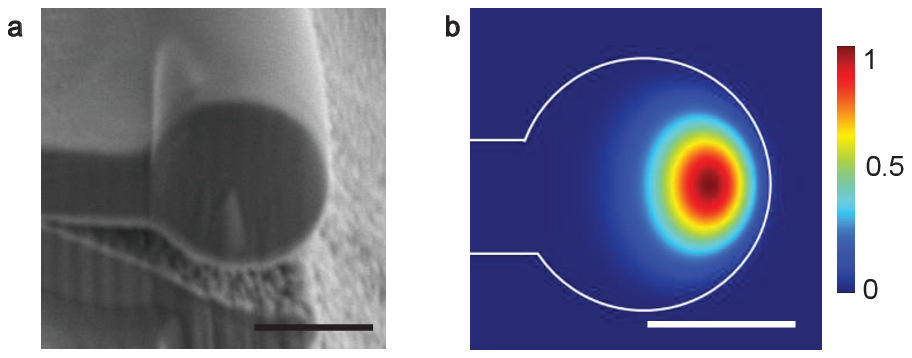}}\caption{(a) SEM
cross-section image of the microtoroid. It is an oblique view taken by an
angle of $56^{\circ}$ with the horizontal direction. (b) COMSOL Multiphysics
finite element method modeling of a fundamental TE mode. The white solid curve
is the boundary of the cavity. The scale bar is 3 $\mathrm{\mu}$\textrm{m}.}%
\label{outer_ring}%
\end{figure}
The field localized at the boundary of the cavity (WGM) is shown in Fig.~\ref{outer_ring}
(toroidal section). It can be seen that the toroid has little influence on the
propagation of chaotic light fields into the disk part of the cavity since the
optical WGMs mainly locate at the same plane with the disk. This observation supports our
two-dimensional model of the microcavity.

\section{Power loss in the silica waveguide bonding with a thick silicon
layer}

The finite-element simulation in Fig.~\ref{power_loss} shows that the power
decrease sharply when the light propagates in the $2$-$\mathrm{\mu}$%
\textrm{m}-thick silica waveguide above a silicon layer with a sufficient
thickness. In particular, the power decrease by $95\%$ when propagating $20$
$\mathrm{\mu}$\textrm{m} (about $19$ wavelengths). The wavelength of the light
used here is $1550$~$\mathrm{nm}$. As for $635$~$\mathrm{nm}$ wavelength,
silicon has strong absorption, which may lead to a faster power
decrease rate.

\begin{figure}[tbh]
\centerline{
\includegraphics[width=14cm,clip]{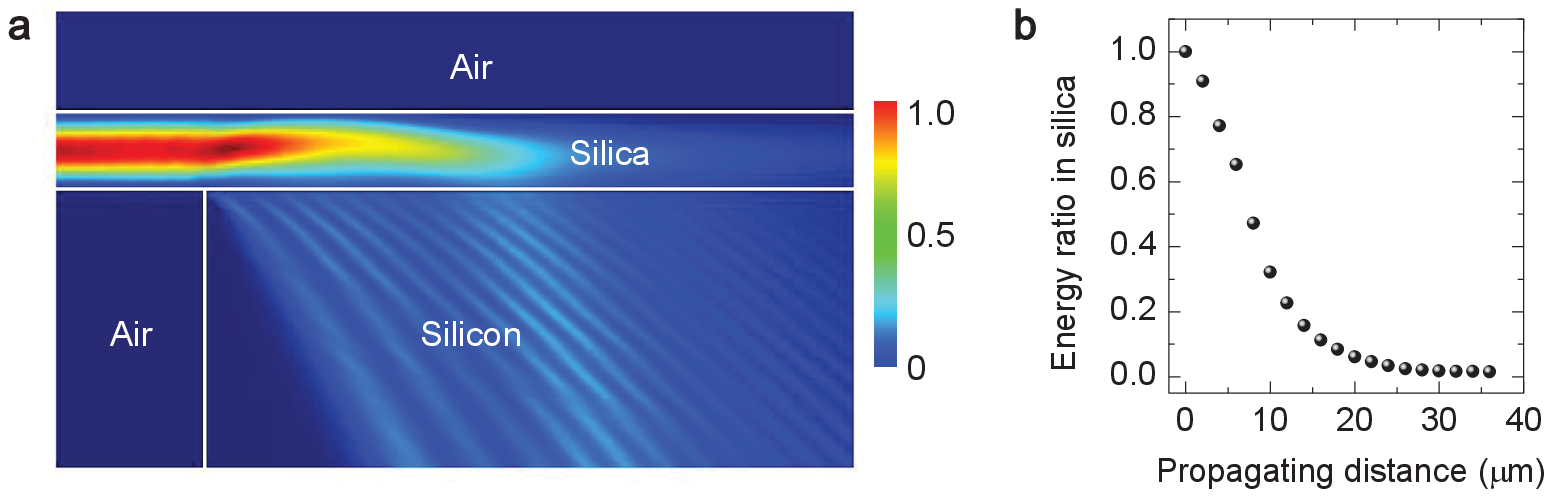}}\caption{(a)
Finite-element-method modeling the light propagating inside the $2$%
-$\mathrm{\mu}$\textrm{m}-thick silica waveguide bonding with a thick silicon
layer. (b) Remaining energy ratio in silica vs. the propagating distance. }%
\label{power_loss}%
\end{figure}

\section{Fiber taper and free-space coupling}

In Fig.~\ref{fiber_taper}(a) we provide an independent cross-check that many
high-$Q$ modes ($Q>10^{5}$) do exist in the deformed microcavity, by exciting
them directly through a tapered fiber. Here the silicon pillar attached to the
microtoroid has largest size, so that dynamical tunneling is inhibited and no
WGM can be excited with the free-space coupling
[Fig.~\ref{fiber_taper}(c)]. By reducing the size of the silicon pillar, as
shown in Fig. 3 in the main text, high-$Q$ modes can be excited indirectly by a
free-space beam.

\begin{figure}[tbh]
\centerline{
\includegraphics[width=.5\textwidth]{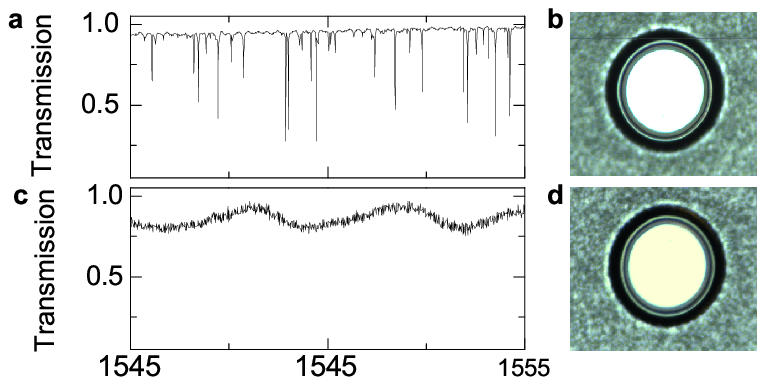}}\caption{ Normalized
transmission and top-view images of the cavity coupled by fiber taper [(a) and
(b)] and free-space laser beam [(c) and (d)]. The absorber-to-cavity ratio
$r\simeq0.83$.}%
\label{fiber_taper}%
\end{figure}

\section{counting modes in transmission spectra}

 From the transmission spectra collected by the photon receiver, we single out the modes that have relatively high $Q$ factors. In particular, we count
 those with $Q>10^5$. There are TE modes and TM modes, which have orthogonal polarization in the microcavity.
 A polarization controller is used to excite modes with one polarization, adjust the polarization to alternatively achieve the highest efficiencies for one kind of modes and suppress the other one. We counted the number of modes in the transmission spectra with both kind of polarizations and add them together to obtain the total number of high$-Q$ regular modes.
The so called TE modes and TM modes are not perfectly orthogonal in the real microcavity, so that some modes may be counted twice, which is the source of the uncertainty in our data.

\section{Ehrenfest time and Lyapunov exponent}
We first give a brief explanation of how the semiclassical correction Eq.~(4) of the main text was obtained in reference [39].
Our first goal is to estimate the fraction of trajectories that escape within the
so called Ehrenfest time $\tau_{Ehr}$, that is the time of quantum-to-classical correspondence.
For an open system, $\tau_{Ehr}$ is the time it takes for a density as large as the size of
the channels to be reduced to one Planck cell by the chaotic motion:
\begin{equation}
\frac{N}{M}e^{-\hat{\mu} t} = \frac{1}{M},
\label{Ehr_find}
\end{equation}
verified for $\tau_{Ehr} = \hat{\mu}^{-1}\log N$ (here $\hat{\mu}$ is proportional to the Lyapunov exponent of the closed
system). Equation~(\ref{Ehr_find}) can also be thought of as
the mean time it takes for a wavepacket the size of a Planck cell to escape, driven by classical
dynamics. We notice $\tau_{Ehr}$ is proportional to the logarithm of the number of channels,
meaning a faster escape implies a longer quantum-to-classical correspondence.
Classically, the probability for an initial area $A_0$ of the phase space to survive
for a time $t$ is
\begin{equation}
P(t) \sim A_0e^{-t/\tau_d},
\label{surv_prob}
\end{equation}
so that the probability for an area to survive in the phase space within Ehrenfest time
is simply $P(\tau_{Ehr}) \sim A_0e^{-\tau_{Ehr}/\tau_d}$.
Now take $A_0=M$, total area of our phase space in units of $h$.
From the definitions we gave of
$\tau_{Ehr}=\hat{\mu}^{-1}\log N$, we can express the survival probability at
Ehrenfest time in terms of the number of open channels and Lyapunov exponent,
as $N^{-1/\hat{\mu}\tau_d}$, and therefore the number of surviving modes as
\begin{equation}
M(\tau_{Ehr}) = MN^{-1/\hat{\mu}\tau_d},
\label{Ehr_svvs}
\end{equation}
as seen in Eq.~(4) of the main text. In some sense, this is the `real' Weyl law here,
since its `fractality' lies in the non-integral power.

The classical estimate of the prefactor $MN^{-1/\tau_d}$ involves Ehrenfest time,
defined for \textit{open} systems as
\begin{equation}
\tau_{Ehr} = \frac{1}{\mu}\log\frac{\tau_H}{\tau_d}.
\label{Ehr_def}
\end{equation}
Here $\mu$ is the Lyapunov exponent of the closed system,
$\tau_d$ is the dwelling time that we are already familiar with, while
$\tau_H$ is the Heisenberg time
\begin{equation}
\tau_H = \frac{h}{\Delta E},
\label{Heis_t}
\end{equation}
with $\Delta E$ mean level spacing, that is average distance (or difference) bewteen consecutive
energy levels. We know, on the other hand, that $E=h\nu$, and we may therefore express
Heisenberg time in terms of the frequency spacing
\begin{equation}
\tau_H = \frac{1}{\Delta\nu},
\end{equation}
and Ehrenfest time as
\begin{equation}
\tau_{Ehr} = \frac{1}{\mu}\log\frac{N}{\Delta\Upsilon}.
\end{equation}
Here $N$ is the number of open channels as we know, whereas $\Delta\Upsilon=MT\Delta\nu$,
that is the mean frequency spacing times the Poincar\'e time (to make it dimensionless),
times the number of states $M$. In plain words, $\Delta\Upsilon$ is the frequency range of
our modes in units of the Poincar\'e time. At this point we can still write
\begin{equation}
\tau_{Ehr} = \frac{1}{\hat{\mu}}\log N,
\end{equation}
provided that
\begin{equation}
\hat{\mu} = \frac{\log N}{\log N - \log\Delta\Upsilon}\mu.
\label{resc_mu}
\end{equation}
Thus we have determined the rescaling to the Lyapunov exponent, following the definition of the
Ehrenfest time.

We may give the rescaling factor an estimate, based on our experimental setup.
Taking for example $T=\frac{a}{2c}~\simeq4\cdot10^{-13}$s ($a$ is the diameter of the cavity,
$c$ the speed of light in the silica), $\lambda=6.35\cdot10^{-7}$m (visible light),
the mean wavelength spacing (from the spectra) as $\Delta\lambda\sim10^{-10}$m, we estimate
$\Delta\nu=\frac{c\Delta\lambda}{\lambda^2}\sim10^{11}$Hz. Moreover, the total number of
states is estimated as $M\simeq Aa/\lambda\simeq40$, with $A$ area of the phase space in exam,
while the mean number of open channels is estimated by the same expression with a smaller area,
indicating the open strip of the phase space. Using the figures above, we get that
\begin{equation}
\hat{\mu}\simeq 1.3\mu.
\end{equation}
%On the other hand, direct numerical simulations in the closed billiard found
%$\mu\in[0.09,0.13]$ (the high uncertainty is due to the sticky regions), and
%therefore the correction due to the opening has not been used.
This correction turns out to be within the uncertainties of the average finite-time
Lyapunov exponent numerically computed for the open billiard, and therefore
it has been neglected.

\section{$\chi^{2}$ test}

The expression for the $\chi^{2}$ test used to assess the goodness of our fits
is
\begin{equation}
\chi^{2}=\frac{1}{\nu}\sum\frac{(x_{\text{\textrm{ob}}}-x_{\text{\textrm{th}}%
})^{2}}{\sigma^{2}},
\end{equation}
$x_{\text{\textrm{ob}}}$ being the observed datum, $x_{\text{\textrm{th}}}$
its theoretical expectation, $\sigma^{2}$ the experimental uncertainty, and
$\nu$ the number of degrees of freedom. In general, $\chi^{2}\gg1$ indicates a
poor model fit, while $\chi^{2}>1$ indicates that the fit has not fully
captured the data (or that the error variance has been underestimated). In
principle, a value of $\chi^{2}=1$ indicates that the extent of the match
between observations and estimates is in accord with the error variance. The
limit of $\chi^{2}<1$ indicates that the model is over-fitting the data:
either the model is improperly fitting noise, or the error variance has been overestimated.

\section{More experimental and fitting results}
In Fig.~\ref{moreexperimentdata} we show more experimental and fitting results.
Fig.~\ref{moreexperimentdata}(a)-(c) are in the infrared wavelength band and (d),(e) are
in the visible wavelength band. All of them support our claim in the main text that
the semiclassical correction (red solid curves) fits the experimental data better than the
purely RMT-based estimate (blue dashed curves), especially at smaller deformation, where
the two predictions differ the most.
\begin{figure}[tbh]
\centerline{
\includegraphics[width=.7\textwidth]{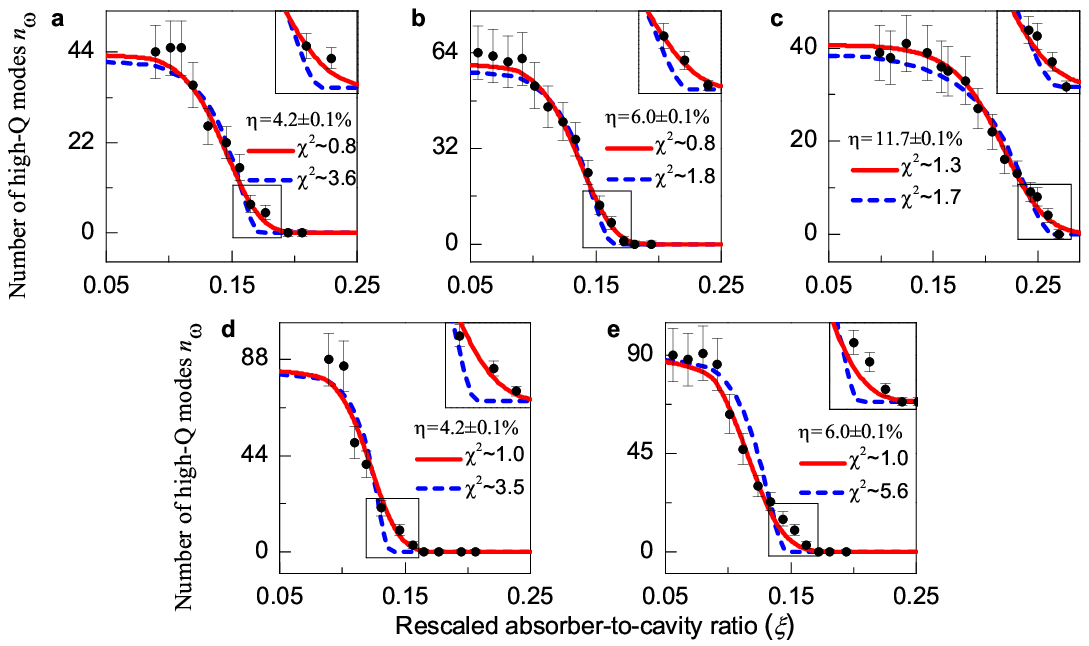}}
\caption{(color online) Number of high-$Q$ regular modes ($n_{\omega}$)
observed in the transmission spectra of the microcavity (Dots), as a function
of rescaled absorber-to-cavity ratio $\xi$. Blue dashed and red solid curves
are respectively RMT- and semiclassical prediction best fits. Here $\sin\theta_{c}\simeq0.69$
and $\sin\theta_{\mathrm{th}}=0.6$. Inset: the area where the two curves
differ most.}%
\label{moreexperimentdata}%
\end{figure}

\section{Table of parameters}
We report the parameters used in our fittings in the following tables
\begin{table} [tbh!]
\centering
\caption[RMT fit]{
Quantities used in relation to
the RMT-based expression (Eq.(6) of
the main text). $\gamma$ is in unit of $T^{-1}$,  with $T\simeq4\cdot10^{-13}$s Poincar\'e time.}
\begin{tabular} {c c c c c c}
\hline\hline
$\Gamma$ &  $\gamma$ &  $\eta$ & $\lambda (\mathrm{nm})$ & $M$ &   $\tilde{\chi}^2$ \\ [0.5ex]
\hline
5.2  &          0.15     &   4.2\%   & 630   &		   40	   &  3.5  \\
3.2 & 	      0.19     &   4.2\%  &  1550  &  		   20      &		3.6  \\
7.2 & 	      0.16     &   6.0\%  &   630  &  		   40 	   &   5.6	   \\
3.6 & 	      0.18     &   6.0\%  &   1550  &  		   20      &   1.8		\\
4.7 &         0.31     &  11.7\%  &   1550  &          25      &   1.7     \\
\hline
\end{tabular}
\label{table:RMT}
\end{table}

\begin{table}[tbh!]
\centering
\caption[Weyl-law fit]{
Quantities used in relation to
the semiclassical prediction (Eq.(7) of
the main text). Both $\gamma$ and $\mu$ are in units of $T^{-1}$,  with $T\simeq4\cdot10^{-13}$s Poincar\'e time.}
\begin{tabular} {c c c c c c c}
\hline\hline
$\Gamma$ &  $\gamma$ &  $\eta$ & $\lambda (\mathrm{nm})$ & $M$ & $\mu$ &   $\tilde{\chi}^2$ \\ [0.5ex]
\hline
2.9  &          0.19     &   4.2\%   & 630   &               40    & 0.13  	  &  1.0  \\
2.0 &         0.23     &   4.2\%  &  1550  &               20      & 0.13        &   0.8  \\
5.0 &         0.20     &   6.0\%  &   630  &                 40      &  0.15	& 1.0        \\
2.5 &         0.21     &   6.0\%  &   1550  &                20         & 0.15		&   0.8              \\
1.5 &         0.38     &  11.7\%  &   1550  &                  25      & 0.21       &   1.3     \\
\hline
\end{tabular}
\label{table:Weyl}
\end{table}

\section{Phase space with a small deformation factor}
The phase space reported in the main text [Fig. 1(b)] belongs to a
microcavity with relatively large deformation factor, $\eta=11.7\%$.
Here we report the classical phase space of the microcavity with the
deformation factor $\eta=4.2\%$ (Fig.~\ref{psos}), together with the survival probability
of a ray in the chaotic region, obtained by analogously to the result shown in
Fig. 5 of the main text. One can see here as well an overall algebraic decay of
the survival probability, and a short-time (transient) hyperbolic behavior.

\begin{figure}[tbh!]
\centerline{
\scalebox{.9}{\includegraphics{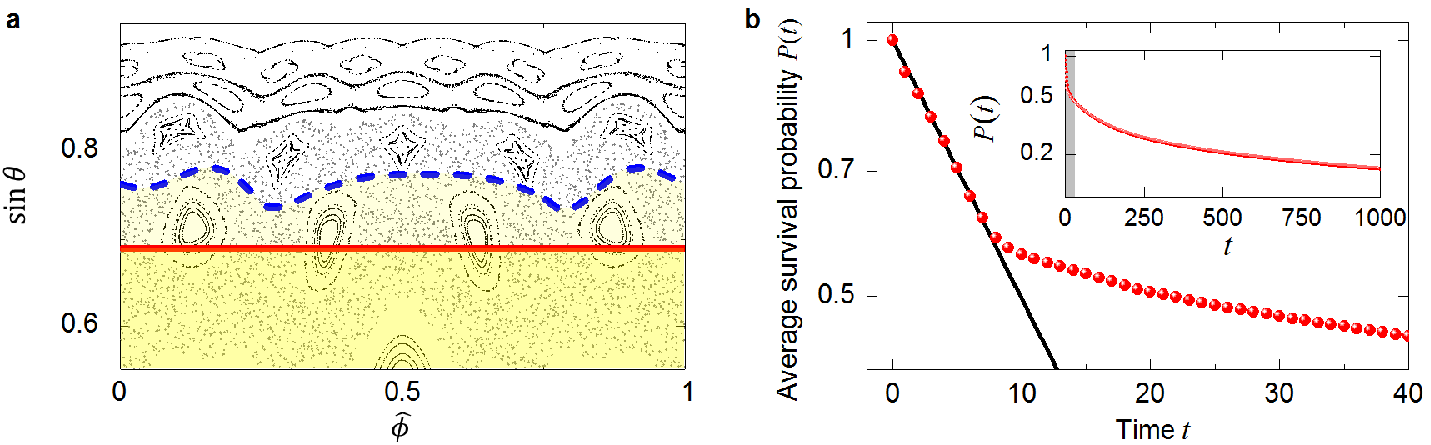}}}
\caption{(a) Poincar\'e surface of section of the microcavity
with deformation factor $\eta = 4.2\%$. Red solid line indicates
the angle of total internal reflection. (b)  (logarithmic scale) Points:
average survival probability $P(t)$ of a ray in the
microcavity
vs. $t$ (in units of Poincar\'e time) at $\xi=0.1$, from $10^{6}$
randomly-started trajectories.
Line: the ansatz $P(t)=\mathrm{exp}(-t/\tau_{d})$, $\tau_{d}=14$. Inset: the
long-time simulation showing algebraic decay.}%
\label{psos}
\end{figure}

\section{Decay rates of WGMs and coupling to chaotic modes}
The approximations
\begin{eqnarray}
\nonumber
\sum_nf_n\frac{V_n}{\gamma_n}
\sim n_\gamma\frac{\bar{f}\bar{V}}{\bar{\gamma}},
\\
\sum_n\frac{V_n^2}{\gamma_n} \sim n_\gamma\frac{\bar{V}^2}{\bar{\gamma}}.
\label{approximations}
\end{eqnarray}
leading to Eq. (3) in the main text for the probability of excitation of a WGM may be thought
of as too rough. In particular, the second expression in~(\ref{approximations})
means that we are ignoring the fluctuations of both the coupling and the decay
of the chaotic modes in consideration. In order to show that we are allowed to do so,
we shall take a step back and rewrite the expression for the amplitude
\begin{equation}
a_\omega =   \frac{E_0\sum_nf_n\frac{V_n}{\gamma_n}}
{\left[\gamma_\omega+i(\omega-\omega_0)\right]+
\sum_n\frac{V_n^2}{\gamma_n}}.
\label{excit_amp}
\end{equation}
Here we recognize the total decay rate of the WGM of frequency $\omega$ as
\begin{equation}
\gamma_\omega^{tot} = \gamma_\omega + \sum_n\frac{V_n^2}{\gamma_n}\simeq
\gamma_\omega + n_\omega\frac{\bar{V}^2}{\bar{\gamma}},
\label{step_back}
\end{equation}
where the first term indicates the intrinsic linewidth of the WGM, while the
second represents the decay into the chaotic modes. That already suggests
that the larger $n$ , the larger $\gamma_{tot}$. If we can really ignore the fluctuations $\gamma_\omega$
in $V$ and $\gamma$, we should be able to see that trend.
The problem is that we do not count $n$ directly, and thus we need to express $\gamma_{tot}$ in terms of some
measurable quantity. A good candidate would be then $n_\omega$, the number of excited WGMs. And
that reminds us of another important approximation:
\begin{equation}
|a_\omega|^2 =   \epsilon^{2}\frac{n_\gamma^2}
{\Gamma^{2}+ n_\gamma^2}.
\label{ex_prob_3par}
\end{equation}
where $\kappa$ is unknown. The assumption is that the number of excited WGMs is
simply proportional to the probability of excitation of one WGM, where, in
reality, $\Gamma = \gamma_\omega\bar{\gamma}/\bar{V}^2$
should be a function of $\omega$, that is even the average coupling
of each regular mode to the chaotic sea depends on where the mode is supported, in the phase space.
It would appear from the literature on dynamical tunneling
[e.g. A. B\"acker et al., Phys. Rev. Lett. 100, 104101 (2008)]
that we are not allowed to  ignore the dependence of $\Gamma$ on $\omega$ as we did, since the couplings $V$
stretch over several orders of magnitude, depending on the regular mode in question.
Still, suppose for a moment we can go on making that approximation. Eq.~(\ref{step_back})
would become, in terms of $n_\omega$,
\begin{equation}
\gamma_\omega^{tot} = \gamma_\omega\left(1+\sqrt{\frac{n_\omega}{\kappa-n_\omega}}\right)
\label{sqrtexp}
\end{equation}
The advantage of this equation is that we have experimental data to fit it to.
Figure~\ref{gamvsn} shows that Eq.~(\ref{sqrtexp})
does qualitatively capture the behavior of the average linewidths of the WGMs, within some errors.
The fitted value for $\gamma_\omega$ corresponds to an average intrinsic $Q$ factor of the order of $10^5$,
which is realistic.
Importantly, the overall enlargement of the average linewidths with the number of observed WGMs constitutes independent evidence for the approximations leading to Eq.~(\ref{ex_prob_3par}) to be reasonable for our experiment.
\begin{figure}[tbh]
\centerline{
\scalebox{.9}{\includegraphics{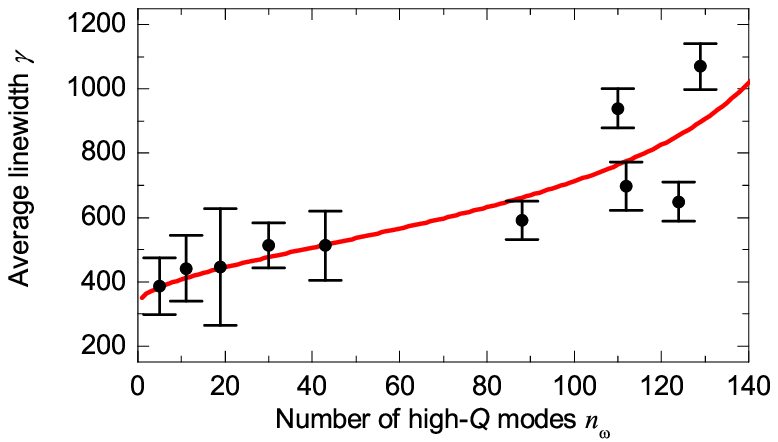}}}
\caption{Points: average linewidth of the excited WGMs vs. their number. Each data
point represents one experiment with a different size of the silicon pillar. As we know, the
number of observed WGMs increases as the size of the pillar decreases. Line: Eq.~(\ref{sqrtexp}),
with $\gamma_\omega = 326$, and $\kappa = 171$.}
\label{gamvsn}
\end{figure}
\end{widetext}


\begin{thebibliography}{99}
\bibitem {ThetaNote}%
In what follows we neglect the dependence of $\theta_a$
on $\phi$ by taking the average value, which is determined
by the ratio $r$ of the mean radius of the absorber to the cavity's.
\bibitem {Note}%
Since we do not have an estimate for the parameter $\epsilon$, all the data sets are
fitted up to a multiplicative constant.
\bibitem {finLyap}
$\mu$ is computed over a short enough time for the
dynamics to be still hyperbolic.
\end{thebibliography}
\end{document}